\documentclass[11pt]{article}

\usepackage{cite}
 \usepackage{subfigure}
\usepackage{multirow}
\usepackage{helvet}
\usepackage{amsmath}
\usepackage{amssymb}
\usepackage{setspace}
\usepackage{graphicx}
\usepackage{empheq}
\usepackage{soul}
\usepackage{tabularx}
\usepackage{array}
\usepackage{float}
\usepackage{braket}
\usepackage[utf8]{inputenc}
\usepackage{url}
\usepackage{hyperref}
\usepackage{slashed}
\usepackage[font=footnotesize,labelfont=bf]{caption}
\usepackage{color}

\def\be{\begin{equation}}
\def\ee{\end{equation}}

\usepackage[a4paper,top=3cm,bottom=3cm,left=2.5cm,right=2.5cm,bindingoffset=0mm]{geometry}

\begin{document}

\begin{center}

{\Large \bf Modelling Coincident Particle Production in\\\vspace*{0.2cm}  Ultraperipheral  Heavy Ion Collisions}

\vspace*{1cm}
L. A. Harland-Lang \\                                          
\vspace*{0.5cm}                                                    
Department of Physics and Astronomy, University College London, London, WC1E 6BT, UK \\   

\begin{abstract}
\noindent In this paper, we present an analysis of coincident particle production in ultraperipheral heavy ion collisions. In particular, we present the first  detailed and differential predictions for coincident $\rho^0$ meson production in association with muon pair production, motivated by the recent ATLAS measurement of this process. These are found to describe the data very well, including the dependence of the coincidence fraction on the ZDC selection and/or other kinematic constraints on the muons. Differential predictions at the level of various kinematic variables are presented, and the calculation is made publicly available in the \texttt{SuperChic} MC generator. We also present general results for coincident two--photon initiated production focussing on muon and electron pair production; while the former is rather suppressed, the latter will be ubiquitous at threshold. The impact of coincident production on exclusivity vetos in ultraperipheral measurements is in addition discussed.

\end{abstract}

\end{center}

\section{Introduction}

In high energy collisions, heavy ions can act as an intense source of initial-state photons, due to their significant electric charges, $Z$. This effect plays a key role in so--called ultraperipheral collisions (UPCs), where the impact parameter separation of the ions is much larger than the range of QCD, and hence no colour flow occurs between the colliding ions. In this environment, photon--initiated (PI) particle production is particularly enhanced, and this has enabled a rich set of experimental measurements at the LHC~\cite{ATLAS:2020epq,ATLAS:2020hii,ATLAS:2022srr,ATLAS:2022ryk,CMS:2022arf,CMS:2020skx,CMS:2024bnt,ATLAS:2025nac} and RHIC~\cite{STAR:2019wlg}.

A particular feature of the large electric charges of the colliding ions is that not just single but multiple photon emission from either ion can play an important role in such processes. Indeed, additional photon exchanges between the colliding ions can quite commonly excite one or both ions into a higher energy state that subsequently decays by emitting a single or multiple neutrons, see e.g.~\cite{Broz:2019kpl,Klein:2020fmr,Harland-Lang:2023ohq}. This effect has been observed in many analyses via zero degree calorimeter (ZDC) detectors, which have been used in UPC measurements at ATLAS~\cite{ATLAS:2020epq,ATLAS:2022srr}, CMS~\cite{CMS:2020skx} and STAR~\cite{STAR:2019wlg}. However, it is equally to be expected that such multiple photon emission can lead to multiple particle production in the central detector. This has been discussed in e.g.~\cite{Hencken:2006ir,Krachkov:2014gba,Klusek-Gawenda:2016suk,vanHameren:2017krz,Zha:2021jhf} in the context of multiple lepton production, but also in~\cite{Klein:1999qj} where the possibility for multiple meson photoproduction processes was considered.

Such coincident particle production has been directly observed for the first time by ATLAS~\cite{ATLAS:2025nac}, where a measurement of $\rho^0$ meson photoproduction coincident with muon pair production has been presented. The fractional rate for this to occur is seen to be at the percent level, consistent with the expectations from~\cite{Klein:1999qj}. However, the results are in addition presented for different ZDC selections and with various kinematic constraints imposed on the final--state muons. This fractional contribution is found to vary with these requirements, as one might expect on general grounds based on the sensitivity of the coincident production rate to the overall peripherality of the interaction, which will itself depend on these constraints. Nonetheless, precise predictions for the overall coincidence rate and this dependence are not so far available in the literature, and hence it is not possible to determine whether these data are indeed quantitatively consistent with expectations based on modelling of the UPC process.

In this paper, we correct this state of affairs, presenting  the first detailed and differential predictions for coincident $\rho^0$ meson production in association with lepton pair production. These predictions are implemented and is made publicly available in the \texttt{SuperChic} Monte Carlo generator. We will focus on the recent ATLAS analysis, and find that indeed these results show a very encouraging level of agreement with theoretical expectations. With this in mind, we  will also provide predictions for further differential observables that may be tested in the future. 

In addition to this, we will discuss the impact any such coincident production may have on observations of UPC production due to individual  photon emission from each ion. Namely, as an exclusivity requirement is imposed in such analyses, a `no--coincident' probability should be accounted for in theoretical predictions (or equivalently corrected for in the data) in a manner that is not necessarily currently performed. The impact of this is expected to be small, at the percent level, but may lead to an improvement in the description of not just the overall rate for certain processes but also the ZDC event fractions.

We will also present broad numerical results for the case of coincident photon--initiated lepton pair production, focussing on the overall expected rates. Consistent with earlier studies, we find that while for multiple muon pair production the expected rates are very low, multiple electron pair production at close to threshold is expected to be ubiquitous. A more detailed analysis will be provided in a future study.

The outline of this paper is as follows. In Section~\ref{sec:cs} we review the basic cross section formulae for UPC production. In Section~\ref{sec:surv} we review how the ion--ion survival factor and mutual ion dissociation are accounted for in this framework. In Section~\ref{sec:coinc} we discuss how coincident particle production is included within this results, for the case of both photoproduction and two--photon initiated production. In Section~\ref{sec:resgam} we present some basic expectations for the case of coincident two--photon initiated lepton pair production. In Section~\ref{sec:resphot} we presented detailed numerical results for the case of coincident $\rho^0$ photoproduction, and compare to the recent ATLAS analysis. Finally, in Section~\ref{sec:conc} we conclude.

\section{Theory}\label{sec:theory}

\subsection{Cross Section Formulae}\label{sec:cs}

The basic theoretical approach follows exactly that outlined in~\cite{Harland-Lang:2023ohq}, see also~\cite{Harland-Lang:2018iur}. Omitting the survival factor for now, the cross section can be written as 
\be\label{eq:csn}
\sigma =\frac{1}{2s}\int {\rm d} x_1 {\rm d}x_2 {\rm d}^2 q_{1\perp}{\rm d}^2 q_{2\perp} {\rm d}\Gamma\frac{1}{\tilde{\beta}} |T(q_{1\perp},q_{2\perp}) |^2\delta^{4}(q_1+q_2-k)\;,
\ee
  where $x_i$ and $q_{i\perp}$ are the photon momentum fractions (see~\cite{Harland-Lang:2019zur} for precise definitions) with respect to the parent ion beams and the photon transverse momenta, respectively.  The photons have momenta $q_{1,2}$, with $q_{1,2}^2 = -Q_{1,2}^2$, and we consider the production of a system of 4--momentum $k = q_1 + q_2 = \sum_{j=1}^N k_j$ of $N$ particles, where ${\rm d}\Gamma = \prod_{j=1}^N {\rm d}^3 k_j / 2 E_j (2\pi)^3$ is the standard phase space volume.  $\tilde{\beta}$ is as defined in~\cite{Harland-Lang:2019zur} and $s$ is the ion--ion squared c.m.s. energy.
  
In \eqref{eq:csn}, $T$ is the process amplitude, and is given by
\be\label{eq:tq1q2}
T(q_{1\perp},q_{2\perp}) = \mathcal{N}_1 \mathcal{N}_2 \,q_{1\perp}^\mu q_{2\perp}^\nu V_{\mu\nu}\;,
\ee  
where $V_{\mu\nu}$ is the $\gamma^*\gamma^* \to X$ vertex, i.e. the amplitude that in the on--shell case would couple to the photon polarization vectors $\epsilon$. The normalization factors are 
\be\label{eq:ni}
\mathcal{N}_i = \frac{2\alpha(Q_i^2)^{1/2}}{ x_i}\frac{F_p(Q_i^2)G_E(Q_i^2)}{Q_i^2}\;.
\ee
where $F_p^2(Q^2)$ is the squared form factor of the ion and 
\be\label{eq:qi}
Q^2_i=\frac{q_{i_\perp}^2+x_i^2 m_{A_i}^2}{1-x_i}\;.
\ee
See~\cite{Harland-Lang:2023ohq} for further details, in particular about the ion form factors.

In the high energy limit, and neglecting the off--shellness of the initial--state photons in the $\gamma\gamma \to X$ process, the above expression reduces to a well known result of the equivalent photon approximation~\cite{Budnev:1975poe}, which is formulated purely at the cross section level. To be precise, we can rewrite \eqref{eq:csn} as 
\be\label{eq:csepa}
{\rm d} \sigma = \int \frac{ {\rm d} x_1 }{x_1} \frac{ {\rm d} x_2 }{x_2}n(x_1) n(x_2) {\rm d}\sigma_{\gamma\gamma \to X}\;,
\ee
in these limits. The flux $n(x_i)$ is given by
\be
n(x_i) =  \int {\rm d}^2 q_{i_\perp}  | {\mathbf N}(x_i,q_{i_\perp})|^2 \;,
\ee
where
\be\label{eq:nibf}
{\mathbf N}(x_i,q_{i_\perp}) =  \frac{ \alpha(Q_i^2)^{1/2}}{\pi} \frac{{\mathbf q}_{i_\perp}}{q_{i_\perp}^2 + x_i^2 m_A^2}F_p(Q_i^2)G_E(Q_i^2)\;.
\ee

\subsection{Survival Factor and Ion Dissociation}\label{sec:surv}

In the high energy limit, and neglecting the off--shellness of the initial--state photons in the $\gamma\gamma \to X$ process,  the cross section~\eqref{eq:csn} can be written as
\be
\sigma \equiv  \frac{1}{2s}\int {\rm d} x_1 {\rm d}x_2 {\rm d}\Gamma\frac{1}{\tilde{\beta}} \delta^{4}(q_1+q_2-k) \,{\rm d} \sigma\;,
\ee
where
\be
{\rm d} \sigma = \int {\rm d}^2 q_{1\perp}{\rm d}^2 q_{2\perp}|T(q_{1\perp},q_{2\perp}) |^2\;.
\ee
Moving to impact parameter space, we can then account for the survival factor, and ion dissociation effects by writing
 \be\label{eq:sigxx}
 {\rm d}\sigma_{X_1 X_2} = \int {\rm d}^2 b_{1\perp}\,{\rm d}^2  b_{2 \perp}\,  |\tilde{T}(b_{1\perp},b_{2\perp})|^2\,\Gamma_{A_1 A_2}(s, b_\perp)\, P_{X_1 X_2}(s,b_\perp)\;.
 \ee
 Here, $\Gamma_{A_1A_2}$ represents the probability that no inelastic scattering occurs at impact parameter $b_\perp=| {\mathbf b}_{1\perp}+{\mathbf b}_{2\perp}|$, and weights the cross section including the survival factor in the appropriate way. It is typically written in terms of the ion--ion opacity $\Omega_{A_1A_2}$ via
\be\label{eq:opac}
\Gamma_{A_1A_2}(s,b_\perp)  \equiv \exp(-\Omega_{A_1 A_2}(s,b_\perp))\;.
\ee
This is given in terms of the opacity due to nucleon--nucleon interactions, $\Omega_{nn}$, which is in turn given by a convolution of the nucleon--nucleon scattering amplitude $A_{nn}$ and the transverse nucleon densities $T_n$, see~\cite{Harland-Lang:2018iur} for a more detailed discussion.

$P_{X_1 X_2}$ is the breakup probability, such that $X_{i} = 0,1,X$ corresponds to the the emission of 0 (i.e. no nuclear excitation) 1, or $X > 0$ neutrons emitted for each ion $i=1,2$. 
 This factorizes into independent breakup probabilities for each ion, i.e.
 \be\label{eq:px1px2}
 P_{X_1 X_2}(b_\perp) = P_{X_1}(b_\perp)P_{X_2}(b_\perp)\;.
 \ee
The lowest order breakup probability for each ion $i=1,2$ is then given by a convolution of the photon emission flux from the ion $j=2,1$ and the $\gamma A \to A^*$ cross section:
\be\label{eq:p1}
P^{1}_{Xn} (b_\perp) = \int\frac{ {\rm d} \omega}{\omega}\, |\tilde{\mathbf N}(x,b_{\perp})|^2  \sigma_{\gamma A \to A^*}(\omega)\;,
\ee
where $\omega$ is the photon energy in the $A$ rest frame, i.e. $\omega=xs/(2m_A)$ in the $s \gg m_A^2$ limit, which holds to very good approximation.  The flux factor $|\tilde{N}|^2$  is given in terms of the Fourier transform
 \be\label{eq:bflux}
 \tilde{\mathbf N}(x,b_{\perp})\equiv\frac{1}{(2\pi)} \int {\rm d}^2 q_{\perp} {\mathbf N}(x,q_{\perp}) e^{-i {\mathbf q}_{\perp}\cdot  {\mathbf b}_{\perp}}\;,
 \ee
where ${\mathbf N}(x_i)$ is defined in \eqref{eq:nibf}.  $\sigma_{\gamma A \to A^*}(\omega)$ is the photon--ion excitation cross section, which has been measured over a wide range of photon energies from fixed target ion scattering experiments, see~\cite{Harland-Lang:2023ohq} for a detailed breakdown of the ingredients which enter this, and which remain the same in the current paper.
 
 \subsection{Including Coincident Production}\label{sec:coinc}
 
 To account for coincident particle production, a very similar approach to that taken for mutual ion dissociation can be taken, see also~\cite{Klein:1999qj,Baur:2001jj,Klusek-Gawenda:2016suk,vanHameren:2017krz,Azevedo:2023vsz} for closely related discussions. For coincident photoproduction, we can in straightforward analogy to~\eqref{eq:p1} write  
 \be\label{eq:pV}
P_{V} (b_\perp) = 2 \int\frac{ {\rm d} \omega}{\omega}\, |\tilde{\mathbf N}(x,b_{\perp})|^2  \sigma_{\gamma A \to V A}(\omega)\;,
\ee
i.e. we simply replace the ion dissociation cross section $\gamma A \to A^*$ with the exclusive photoproduction cross section $\gamma A \to A V$ of a state V. This can be calculated, or evaluated from the relevant direct experimental determinations of the photoproduction process. The factor of 2 here accounts for the fact that the coincident production may be due to photon emission from either ion.
 
To account for the kinematic dependence of this coincident  production we then as usual move back to transverse momentum space. That is, at the amplitude level we should replace the $T(q_{1\perp},q_{2\perp})$ entering \eqref{eq:csn} with
\be\label{eq:tqts2}
T_{{\rm S^2}, \, X_1X_2}^V(q_{1\perp},q_{2\perp}) =\frac{1}{(2\pi)^2}\int {\rm d}^2 b_{1\perp}{\rm d}^2 b_{2\perp}\,e^{i {\mathbf q}_{1\perp}\cdot {\mathbf b}_{1\perp}}e^{-i {\mathbf q}_{2\perp}\cdot {\mathbf b}_{2\perp}}\tilde{T}(b_{1\perp},b_{2\perp})\hat{\Gamma}_{X_1X_2}^V(s,b_\perp)^{1/2}\;,
\ee
where the `$S^2$' indicates that the ion--ion survival factor is now appropriately accounted for, and the $X_1, X_2$ ($V$) indicate whether ion dissociation (coincident production) occur. Here  we have defined
\be\label{eq:gamprob}
\hat{\Gamma}_{X_1X_2}^V(s,b_\perp)^{1/2}= \left[ \Gamma_{A_1A_2}(s,b_\perp) P_{X_1 X_2}(b_\perp) P_{V} (b_\perp)\right]^{1/2}\;.
\ee
We have dropped the ($A_1 A_2$) dependence on the ion type on the left hand side, which is implied, as it is for the ion dissociation and coincident photoproduction probabilities. We also note that exactly the same expression is used, but with the $P_V$ omitted, for the case with no coincident production.

A convenient form for this comes from defining
\be\label{eq:pmc1}
\mathcal{P}_{X_1X_2}^V(s,k_\perp)\equiv \frac{1}{(2\pi)^2}\int {\rm d}^2 b_{\perp}\,e^{i {\mathbf k}_{\perp}\cdot {\mathbf b}_{\perp}}\hat{\Gamma}_{X_1X_2}^V(s,b_\perp)^{1/2}\;,
\ee
in terms of which we have
\be
T_{{\rm S^2}, \,X_1X_2}^V(q_{1\perp},q_{2\perp}) = \int  {\rm d}^2 k_{\perp}\, T(q_{1\perp}',q_{2\perp}')\,\mathcal{P}_{X_1X_2}^V(s,k_\perp)\;,
\ee
where $q_{1\perp}' =q_{1_\perp} - k_\perp$ and $q_{2\perp}' = q_{2\perp} + k_\perp$. 

\begin{figure}
\begin{center}
\includegraphics[scale=0.62]{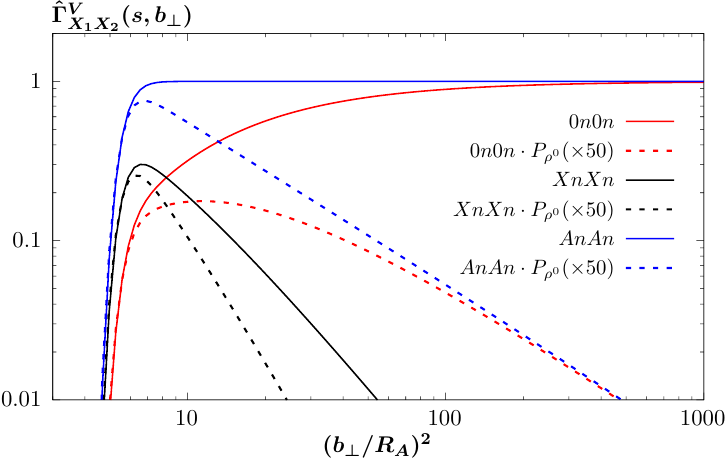}
\includegraphics[scale=0.62]{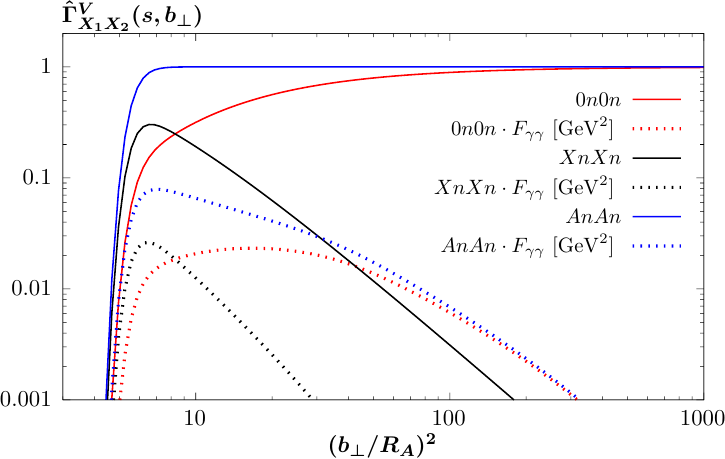}
\caption{\sf Breakup probabilities for no or multiple neutron emission at $\sqrt{s_{nn}}=5.02$ TeV in PbPb collisions, as function of the squared ion--ion impact parameter. The solid curves correspond to the exclusive case. (Left) Dashed curves include the probability of coincident $\rho^0$ production, multiplied by a factor of multiplied by a factor of 50 for ease of comparison. (Right) Dotted curves include the flux contribution to  coincident photon--initiated production \eqref{eq:fgam}, as described in the text. Results are multiplied by the ion--ion survival probability in all cases. }
\label{fig:prob}
\end{center}
\end{figure}

To assess the expected contribution from coincident photoproduction we note that, as discussed in~\cite{Harland-Lang:2023ohq}, in the $b_\perp > R_A$ region (i.e. treating the ion as a point--like charge) and for moderate $b_\perp$ values, the photon flux \eqref{eq:bflux} behaves as
 \be\label{eq:napprox}
|\tilde{\mathbf N}(x,b_{\perp})|^2 \approx \frac{Z^2\alpha}{\pi^2} \frac{1}{b_\perp^2}\;.
\ee
That is, it is peaked towards low values of the ion--ion impact parameter, $b_\perp$, and hence will be favoured for those production processes that favour lower values of this. More precisely, at $b_\perp = 2 R_A$ we have
 \be\label{eq:napprox_2ra}
|\tilde{\mathbf N}(x,2 R_A)|^2 \approx 0.003 \, \left(\frac{Z}{Z_{\rm Pb}}\right)^2 \, {\rm mb}^{-1}\;,
\ee
which represents the minimum relevant suppression with impact parameter, with this falling like $1/b_\perp^2$ at higher values. Therefore, for PbPb collisions and a photoproduction cross section that is $O({\rm mb})$, the maximum relevant coincidence probability will be at the percent level, as discussed in~\cite{Klein:1999qj}.

To illustrate the trend with impact parameter, in Fig.~\ref{fig:prob} (left) we plot the square of \eqref{eq:pmc1}, i.e. the  breakup probabilities for no or multiple neutron emission at $\sqrt{s_{nn}}=5.02$ TeV in PbPb collisions, as a function of the squared ion--ion impact parameter, including the ion--ion survival factor and with/without (shown by the dashed/solid lines) coincident $\rho^0$ meson production. The precise method for calculating the $\rho^0$ photoprodution cross section is described in the following section, and the rate is multiplied by a factor of 50 in the plot for ease of comparison. We note that here $Xn$ corresponds to 1 or more neutrons being produced, while $An$ is inclusive with respect to neutron production.

Our basic expectation will then be that the relative fraction of the integrand that remains after requiring coincident $\rho^0$ production will determine the relative rates at which this occurs between the different dissociation cases. In particular, for the $0n0n$ case, which is most peaked towards higher $b_\perp$, we will expect the smallest rate, and for the $XnXn$ case, which is most peaked towards lower $b_\perp$, we will expect the largest rate. For other cases, such as $AnAn$, which lie in between, we will likewise expect the coincidence rate to be between these. As we will see, this is indeed observed in the quantitative predictions.

We note that, similarly to some cases without coincident production discussed in~\cite{Harland-Lang:2023ohq}, we sometimes have to deal with the fact that the high $b_\perp$ behaviour of the integrand in \eqref{eq:pmc1}:
\be
I_{X_1X_2}^V(b_\perp)=b_\perp J_0(b_\perp k_\perp) \left[P_{X_1 X_2}(b_\perp) P_{V} (b_\perp)\right]^{1/2}\;,
\ee
while strictly speaking  convergent, requires some manipulation for this to be achieved numerically. This in particular occurs when the term inside the squares brackets scales as $\sim 1/b_\perp^2$, which from \eqref{eq:napprox} we can see will occur when $P_{X_1 X_2}(b_\perp)\sim $ const., i.e. for the $AnAn$ and $0n0n$ cases. This is dealt with in precisely the same way as in~\cite{Harland-Lang:2023ohq}, namely by dividing the integral into a piece which can be integrated analytically and a piece that falls more steeply with $b_\perp$ and can therefore be safely integrated numerically. We refer the reader to this reference for further details, which also describes how the cases without coincident production are dealt with.

Finally, if instead of coincident photoproduction we are interested in coincident $\gamma \gamma \to X$ production, then this can be accounted for in a rather similar way. In particular, the photoproduction cross section is simply given by integrating the probability \eqref{eq:pV} over the ion--ion impact parameter, i.e. we have
\be\label{eq:pvb}
 \sigma_{A_1 A_2 \to A_1 A_2 X} =   \int {\rm d}^2 b_{\perp}P_{X} (b_\perp)\;.
\ee
For the purely photon initiated case we can simply rewrite \eqref{eq:csepa} in impact parameter space 
\be
 \sigma_{A_1 A_2 \to A_1 A_2 X}^{\gamma\gamma} =  \int {\rm d}^2 b_{1\perp}{\rm d}^2 b_{2 \perp} \int \frac{ {\rm d} x_1 }{x_1} \frac{ {\rm d} x_2 }{x_2} |\tilde{\mathbf N}(x_1,b_{1\perp})|^2  |\tilde{\mathbf N}(x_2,b_{2\perp})|^2\,\sigma_{\gamma\gamma \to X}\;,
\ee
where the arguments of the $\gamma \gamma \to X$ cross section are omitted for brevity. Here, as above, the $b_{i\perp}$ are the impact parameters between the ion $i$ and the produced system $X$. From this we then write
\be\label{eq:pv2gam}
P_{X, \gamma\gamma}(b_\perp) = \frac{1}{2} \int {\rm d}^2 b_{\perp}'  \frac{ {\rm d} x_1 }{x_1} \frac{ {\rm d} x_2 }{x_2} |\tilde{\mathbf N}(x_1,b_{1\perp})|^2  |\tilde{\mathbf N}(x_2,b_{2\perp})|^2\,\sigma_{\gamma\gamma \to X}({\mathbf b}_{1\perp},{\mathbf b}_{2,\perp})\;,
\ee
where ${\mathbf b}_{1,2 \perp}=({\mathbf b}_\perp \pm {\mathbf b}_\perp')/2$. This can then be suitably inserted into \eqref{eq:pmc1}, with an appropriate symmetry factor if the coincident and original states produced are identical, and the phase space integral for the primary production overlaps with that of the secondary production. A similar discussion is presented in~\cite{Klusek-Gawenda:2016suk,vanHameren:2017krz,Azevedo:2023vsz}, although here the impact parameter dependence of the subprocess cross section itself is omitted, which is in principle important. 

\section{Results}

\subsection{Coincident two--photon initiated production: general results}\label{sec:resgam}

From \eqref{eq:pv2gam}, we can make some general comments about the probability for coincident photon--initiated dilepton production. Making the approximation that we can drop the impact parameter dependence of the subprocess cross section as well as, according to \eqref{eq:napprox}, the $x$ dependence of the photon fluxes, we are interested in
\be
P_{X, \gamma\gamma}(b_\perp) = F_{\gamma\gamma}(b_\perp) \cdot \sigma_{X, \gamma\gamma}\;,
\ee
where we have defined
\begin{align}\label{eq:fgam}
 F_{\gamma\gamma}(b_\perp) &=   \frac{1}{2}\int {\rm d}^2 b_{\perp}'  |\tilde{\mathbf N}(b_{1\perp})|^2  |\tilde{\mathbf N}(b_{2\perp})|^2\;,\\
 \sigma_{X, \gamma\gamma} &=\int  \frac{ {\rm d} x_1 }{x_1}  \frac{ {\rm d} x_2 }{x_2} \,\sigma_{\gamma\gamma \to X}\;,
\end{align}
suitably dropping arguments as per the approximation above. For the latter object we have 
\be\label{eq:siggamll}
\sigma_{\gamma\gamma \to ll} \sim \frac{2\pi \alpha^2}{M_{ll}^2}\;,
\ee
see e.g. Appendix A of ~\cite{Harland-Lang:2011mlc}, where we take the integral over the dilepton angular distribution, within suitable experimental cuts, to be $\sim 1$. Integrating over a central unit of rapidity $|y_{ll}|<0.5$ and from a threshold $M_{ll}^{\rm min}$ we have
\be\label{eq:vgam}
 \sigma_{X, \gamma\gamma} \sim \frac{8\pi \alpha^2}{(M_{ll}^{\rm min})^2} \sim \frac{10^{-3}}{(M_{ll}^{\rm min})^2} \;.
\ee
Turning to the former object, \eqref{eq:fgam}, in Fig.~\ref{fig:prob} (right) we  show the result for the breakup probability as defined in \eqref{eq:gamprob}, but including this flux contribution. We can see that the coincidence probability is, similarly to the photoproduction case, peaked at lower ion--ion impact parameters. Therefore, qualitatively similar conclusions about the kinematic and ZDC class dependence of the coincidence probability are expected to hold, although not quantitatively, as the scaling with impact parameter is not identical. 

In terms of the level of suppression, we find that that the flux component of the coincidence probability is enhanced by roughly two orders of magnitude relative to the photoproduction case, primarily due to the increased $\sim Z^4$ scaling of the result in comparison to the $\sim Z^2$ scaling of the photoproduction case. That is, at $b_\perp = 2 R_A$ we have
\be
 F_{\gamma\gamma}(b_\perp) \approx 0.3 \left(\frac{Z}{Z_{\rm Pb}}\right)^4 \, {\rm mb}^{-1}\;,
\ee
i.e. roughly 0.1 in units of ${\rm GeV}^2$ for PbPb collisions, as observed  in Fig.~\ref{fig:prob} (right).

Multiplying this by the above approximate expression \eqref{eq:vgam} gives the full coincidence probability, and we can see leads to a significant, $\sim 10^{-6}$ $(10^{-4}$), suppression at a dilepton invariant mass of 10 (1) GeV, with respect to the solid curves, where no coincident production is required. Therefore, for reasonable values of the dilepton invariant mass the coincidence rate is indeed expected to be very low, broadly consistent with the results of~\cite{Klusek-Gawenda:2016suk,vanHameren:2017krz}. However the above results are clearly only an estimate, and moreover rather a conservative one, as the contribution from the angular integral in \eqref{eq:siggamll} is more precisely rather larger than unity, while integrating over a wider angular and rapidity region will increase this further. If the kinematic cuts on the additional lepton pair are sufficiently loose this will increase significantly, by up to an order of magnitude. 

In the extreme, if we integrate down to threshold (and again over unit central rapidity), we have
\be
 \sigma_{X, \gamma\gamma} \sim \frac{\pi\alpha^2}{m_l^2} \sim 0.02 \,(600)\; {\rm GeV}^{-2}\;,
\ee
for muon (electron) pair production. That is, after accounting for the photon flux suppression as in Fig.~\ref{fig:prob} (right), the coincidence rate for muon pair production would be at the permille level, while for electron pair production this will become  significantly larger than one, and will remain so even beyond threshold. That is, while coincident muon pair production appears to rather challenging to observe, for coincident electron production we expect multiple pair production to occur for essentially all UPC collisions. This can be readily dealt with in a similar manner to the case of mutual ion dissociation, as discussed in~\cite{Harland-Lang:2023ohq}, i.e. through a standard process of unitarisation. The key issue is, as discussed in~\cite{Harland-Lang:2021ysd}, the extent to which such production near the electron mass threshold will be observable, and/or will impact on any exclusivity veto, or rather how far above threshold the production must be for this to be the case. We note that the above estimate for muon pair production is entirely consistent with the calculation of~\cite{Hencken:2006ir}, which predicts a contribution of below 1\% for multiple muon pair production at threshold, see~\cite{Harland-Lang:2021ysd} for more discussion and~\cite{Krachkov:2014gba,Zha:2021jhf} for other studies.

We will focus for the rest of this paper on the photoproduction case, leaving a more detailed study of coincident two--photon production for future study, but here we note that again the rate of coincident production is highly sensitive to the ion--ion impact parameter, and hence we will expect similar (though not identical) dependencies on the final--state kinematics and ion dissociation selection.

\subsection{Coincident photoproduction: comparison to ATLAS data}\label{sec:resphot}

\begin{figure}
\begin{center}
\includegraphics[scale=0.7]{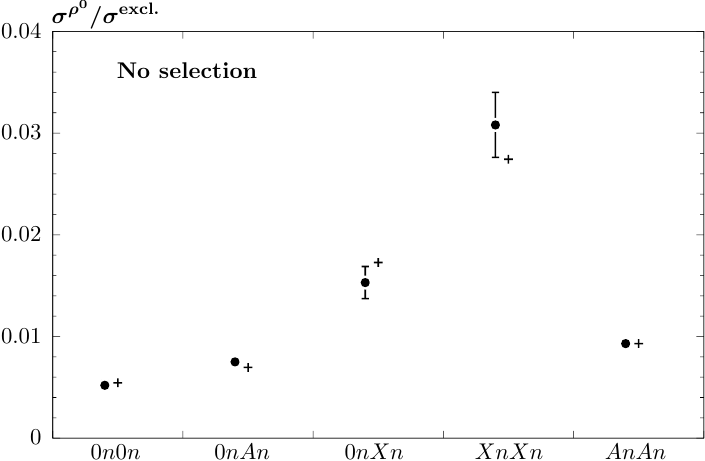}
\caption{\sf The fraction of $\mu^+ \mu^-$ UPC events with coincident $\rho^0$ production for different ZDC selections. The left solid points correspond to the ATLAS data~\cite{ATLAS:2025nac} while the right crosses correspond to the SC predictions, as described in the text. For the data the systematic and statistical uncertainties are added in quadrature, and are in some cases not visible on the plots due to their small size.}
\label{fig:ATLAScomp_nosel}
\end{center}
\end{figure}

In this section we will present numerical results for the case of coincident photon--initiated muon pair and $\rho^0$ meson photoproduction, motivated by the recent ATLAS measurement of this process~\cite{ATLAS:2025nac}. This is calculated using the approach outlined in the previous sections, and is implemented in the \texttt{SuperChic} MC generator. This simulates the differential cross section for dimuon production with coincident $\rho^0$ photoproduction, for standard ZDC selections, but does not explicitly generate the $\rho^0$ meson final--state and its decay products. In particular, the kinematics of the produced $\rho^0$, and its decay, can be simulated if necessary (e.g. to evaluate acceptance corrections and so on) by a dedicated MC generator for $\rho^0$ photoproduction, such as  \texttt{STARLight}~\cite{Klein:2016yzr}.

We set the normalization of the $\gamma A \to \rho^0 A$ subprocess cross section in \eqref{eq:pV} using the experimental value from the ALICE measurement~\cite{ALICE:2020ugp} of $\rho^0$ production in UPCs at 5.02 TeV. In particular, as this is expected to depend rather weakly on the photon energy~\cite{Klein:1999qj,Frankfurt:2015cwa} we simply assume a constant cross section. Applying \eqref{eq:pvb} we can then translate this into a UPC cross section; we find a value of 2.3 mb for $\sigma_{\gamma A \to V A}$ gives a cross section in the ALICE case of 560 mb in the central $|y_{\rho^0}|<0.2$ bin, in excellent agreement with the experimental value of  537 $\pm$ 4.6 (stat) ${}^{+46.1}_{-42.0}$ (sys) mb. 
This is broadly consistent with other determinations, see e.g.~\cite{Frankfurt:2015cwa}.  We note that the precise value is chosen so as to match the ATLAS $AnAn$ (no selection) results, subject to the considerations below, but as noted above, it is entirely consistent with the ALICE data.  Taking this, we then integrate over a photon energy range roughly in line with the ATLAS rapidity selection, namely for $|y_{\rho^0}|<2.5$. While this is our baseline, as the coincident cross section is simply linearly dependent on the overall normalization of  $\sigma_{\gamma A \to V A}(\omega)$, it is straightforward to vary this when comparing to data, if required.

In more detail, in the ATLAS analysis the data for $\rho^0 \to \pi^+ \pi^-$ in fact also includes a contribution from coherent non--resonant $\pi^+\pi^-$ photoproduction. This is subtracted in the ALICE data, but not in the ATLAS case, and hence we would expect including  coherent $\rho^0$ photoproduction alone, with a normalization extracted from the ALICE data,  to undershoot the ATLAS measurement of the coincident fractions. A fit to the dipion mass spectrum, including the non--resonant piece, is provided by numerous experiments~\cite{ZEUS:1997rof,ALICE:2015nbw,STAR:2017enh,LHCb:2025fzk}, and integrating over this we find a relative contribution of $\sim 15\% - 35\%$ from non--resonant production, depending on the fit/dataset. In general we will expect the precise contribution to depend on the particular beam energies (and types) as well the event selection, and we note that the above hadron--hadron results would in principle require correcting to the $\gamma A$ level for our purposes (even if the mass dependence of the photon flux from the ion will be relatively flat over this limited mass region). Taking the \texttt{STARLight}~\cite{Klein:2016yzr} prediction for the ATLAS event selection, we find a value of $\sim 20\%$\footnote{To be precise to arrive at this the code was suitably modified such that the cross section normalization is tied to the integral over the resonant $\rho^0$ Breit--Wigner alone, i.e. so that the non--resonant contribution is not absorbed into this normalization.}, although the precise number will depend on the underlying fit used, in this case to ZEUS data~\cite{ZEUS:1997rof}. Hence, this can only be considered as an estimate, but nonetheless we should increase our effective  $\rho^0 \to \pi^+ \pi^-$ by roughly such a factor.

In addition to the above, while the predictions correspond to  $|y_{\rho^0}|<2.5$, in the ATLAS analysis results are presented with a somewhat different requirement. Namely, no acceptance correction due to the requirement that the decay pions must lie within the fiducial region ($|\eta_\pi|<$ 2.5, $p_{\perp, \pi}>$ 100 MeV) is applied. In~\cite{ATLAS:2025nac} it is commented that roughly 76\% of $\rho^0$ events with $|y_{\rho^0}|<2.5$ are found in the ATLAS event  selection on the pions from the $\rho^0$ decay, when using the \texttt{STARLight} MC generator~\cite{Klein:2016yzr}. In other words, to match the ATLAS selection we should scale our effective value of $\sigma_{\gamma A \to V A}$ by $\sim 0.76$. Combining this decrease with the increase described above due to non--resonant production, we find these to first approximation cancel one another. That is,  for the results which follow we can simply apply a prediction for purely resonant $\rho^0$ photoproduction, with normalization tied to the ALICE data, and consider that the effect of the pion acceptance and non--resonant production have opposite effects. For future data, a subtraction of the non--resonant contribution would enable a more precise comparison. We also emphasise again that a differing acceptance factor and/or impact of non--resonant $\pi^+\pi^-$ production can be straightforwardly accounted for by simply scaling all coincident cross sections by the relevant factor.

\begin{figure}[t]
\begin{center}
\includegraphics[scale=0.6]{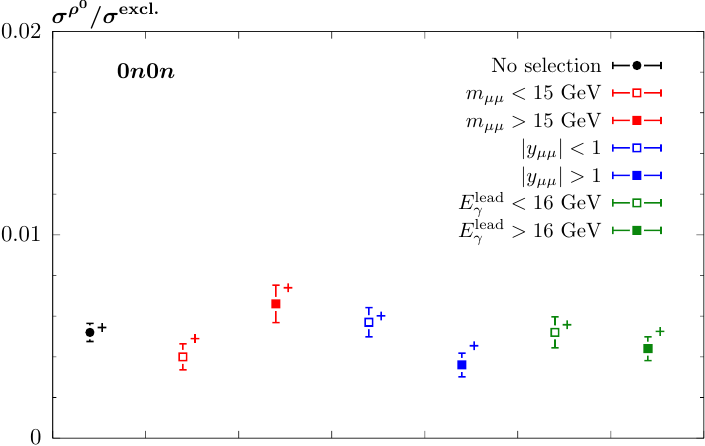}
\includegraphics[scale=0.6]{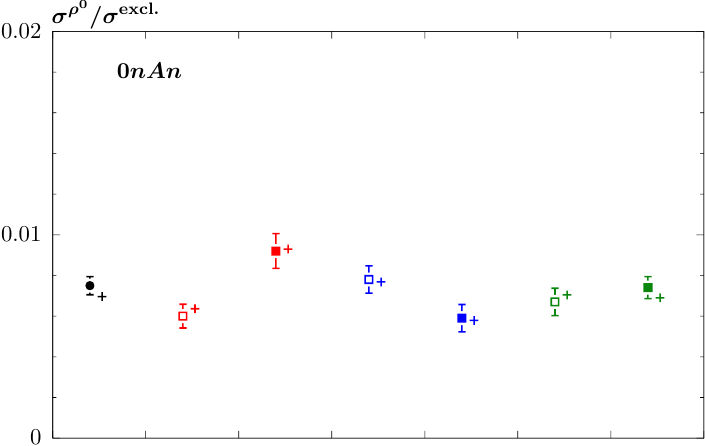}
\includegraphics[scale=0.6]{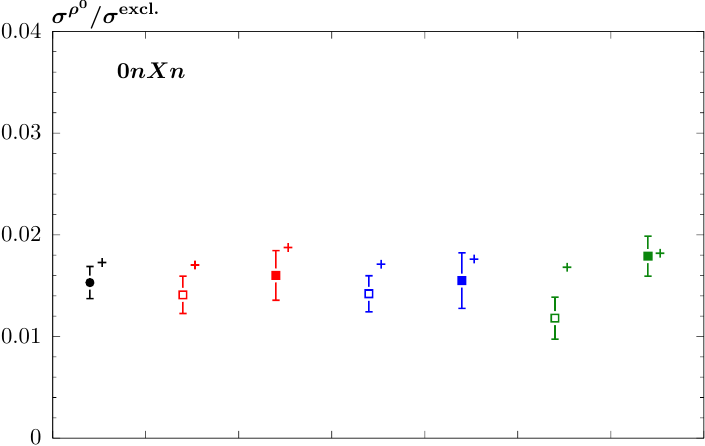}
\includegraphics[scale=0.6]{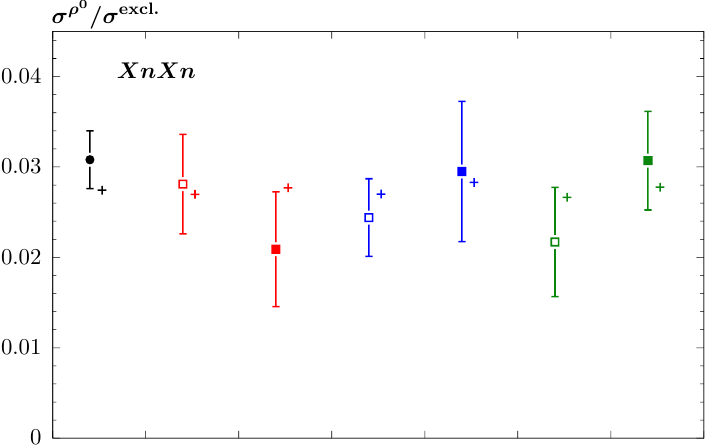}
\includegraphics[scale=0.6]{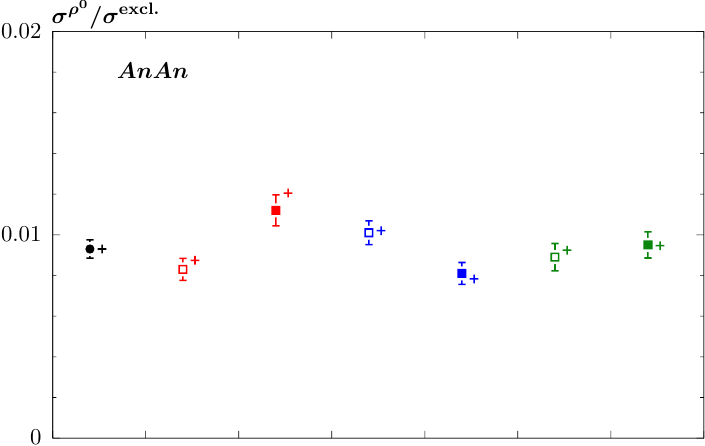}
\caption{\sf As in Fig.~\ref{fig:ATLAScomp_nosel} but also showing the results with different kinematic selections applied to the produced muons. The different ZDC selections are shown in separate plots, as indicated.}
\label{fig:ATLAScomp_sel}
\end{center}
\end{figure}

We first consider the overall fraction of $\mu^+ \mu^-$ UPC events with coincident $\rho^0$ production for different ZDC selections, with no additional kinematic requirements applied. All theoretical results are calculated at the higher c.m.s. energy value of 5.36 TeV, where most of the data are taken; the correction to account for the data at lower energy can be safely ignored here. This is shown in Fig.~\ref{fig:ATLAScomp_nosel}, where a comparison is presented to the ATLAS data~\cite{ATLAS:2025nac}. The overall trend for an increasing fraction from the $0n0n$ to the $XnXn$ ZDC selections, as discussed in the previous section, is indeed observed in the predictions, with a sizeable difference seen between these two most extreme cases. Moreover, the normalization of the coincident fractions is matched well.

\begin{figure}[t]
\begin{center}
\includegraphics[scale=0.63]{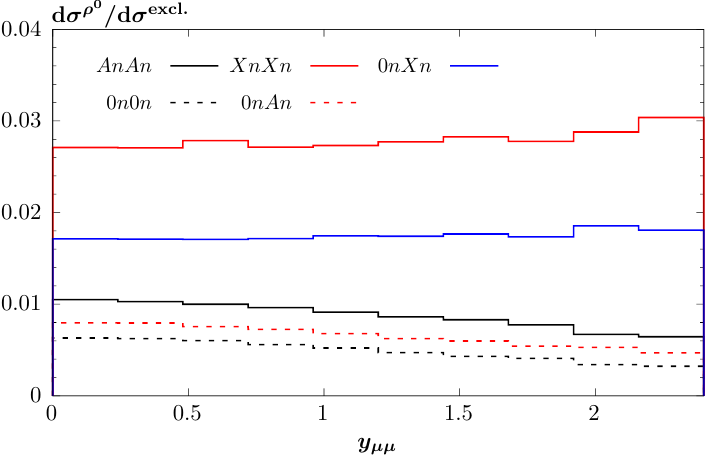}
\includegraphics[scale=0.63]{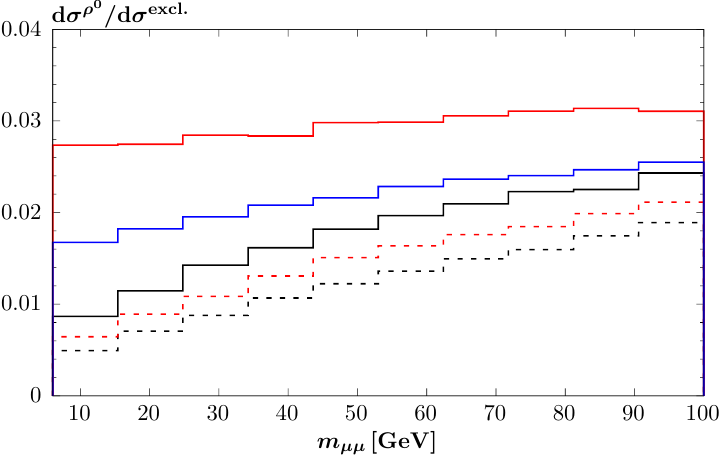}
\includegraphics[scale=0.63]{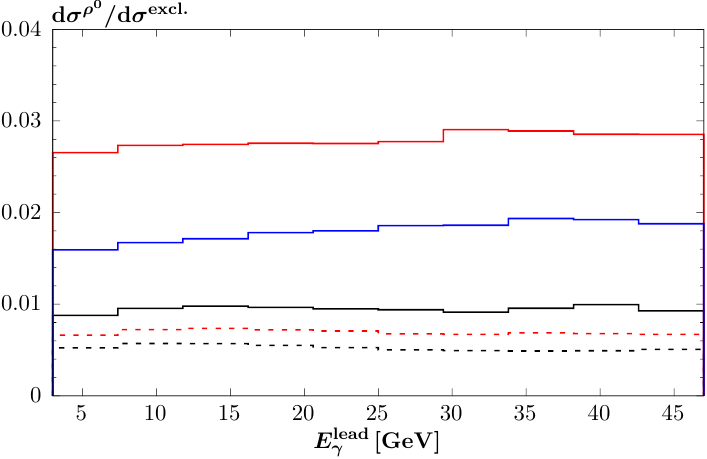}
\caption{\sf The predicted differential fractional distributions for $\mu^+ \mu^-$ UPC events with coincident $\rho^0$ production and different ZDC selections. Results correspond to the same event selection as in Fig~\ref{fig:ATLAScomp_sel}, and for are given with respect to the kinematic variables for which the selections applied in this figure, namely the dimuon invariant mass, rapidity and the leading initial--state photon energy.}
\label{fig:dists}
\end{center}
\end{figure}

We next turn to the results with additional kinematic requirements imposed on the final--state muons, namely requiring the dimuon rapidity, invariant mass or maximum initial--state photon energy to lie above or below a given threshold; the latter is reconstructed from the dimuon kinematics via $E_\gamma = \frac{m_{\mu\mu}}{2}e^{\pm y_{\mu\mu}}$. These are shown in Fig.~\ref{fig:ATLAScomp_sel}, with each panel corresponding to a particular ZDC selection. A noticeable dependence on these kinematic requirements is observed in the results, and in the data, most clearly for the invariant mass and rapidity selections, where a larger $m_{\mu\mu}$ (lower $y_{\mu\mu}$) leads in general to an increased coincidence fraction. These trends are matched in the data, with the overall agreement being rather good.

To examine the reason for this dependence we show in Fig.~\ref{fig:ATLAScomp_sel} the corresponding predicted kinematic distributions for these variables. As commented on in~\cite{ATLAS:2025nac}, the key factor is the changing peripherality of the interaction with these selections. Indeed the overall trends are precisely as discussed in~\cite{Harland-Lang:2021ysd} for the case of the ion--ion survival factor. In particular, by requiring a larger $m_{\mu\mu}$ the photon momentum fractions are increased, which leads a larger average photon virtuality, $Q^2$, and hence a less peripheral interaction. Recalling Fig.~\ref{fig:prob} and the discussion accompanying it, such lower impact parameters will favour coincident production, as well as leading to a lower ion--ion survival factor, as discussed in~\cite{Harland-Lang:2021ysd}. For the rapidity selection, the general expectation is less clear, as an increase in rapidity leads to an increased photon momentum fraction in one case, and a decrease for the other. However, the trend for a decreased coincidence rate with increasing rapidity for the $AnAn$, $0n0n$ and $0nAn$ selections, is again consistent with the prediction of an increased survival factor (i.e. an overall less peripheral interaction) in this case. In the case of the maximum photon energy, for which a larger energy corresponds to a larger dimuon invariant mass and a larger rapidity, these two counteracting effects tend to cancel out, and the predicted dependence is rather weak, though not completely flat.

Looking at the difference between the ZDC selections in more detail, we can see that the trend for an increased coincidence rate for increasing dimuon invariant mass is strongest for the cases such as $0n0n$. This is as we might expect, given these tend to be the most peripheral interactions, and hence one might expect an increased sensitivity to a reduction in the impact parameter. Conversely, for the $XnXn$ selection, where the interaction is relatively less peripheral, there is less scope for reducing the impact parameter and increasing the coincidence rate further. For the dimuon rapidity, as discussed above the general expectation is not clear and therefore more dependent on the precise process. Indeed, we can see that the overall trend for the $XnXn$ and $0nXn$ selections is rather flat, with no particular decrease in coincidence rate at larger rapidities. Given this, there is some mild trend for a increase in coincidence rate for these selections, with respect to the maximum photon energy, driven by the increase with dimuon invariant mass.

\begin{figure}
\begin{center}
\includegraphics[scale=0.6]{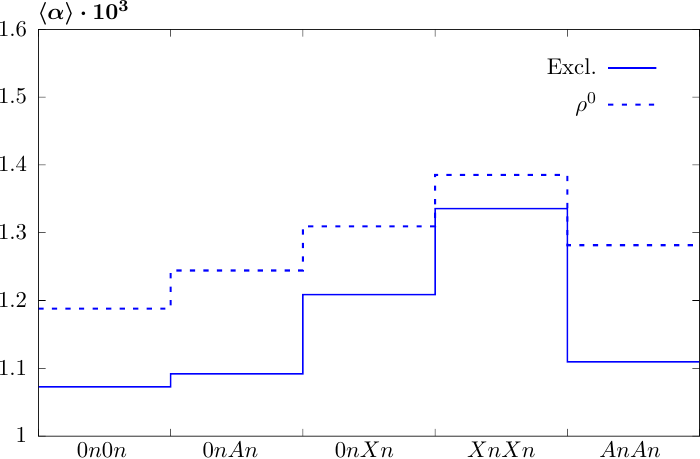}
\caption{\sf Predicted average dimuon acoplanarity for the same event selection as in the ATLAS analysis~\cite{ATLAS:2025nac}. Results are shown for the purely exclusive process (solid) and with coincident $\rho^0$ production.
}
\label{fig:aco}
\end{center}
\end{figure}

In Fig.~\ref{fig:aco} we show the dependence on an additional kinematic variable not considered in the ATLAS analysis, namely the dimuon acoplanrity $\alpha= 1-\Delta\phi_{\mu\mu}/\pi$. We in particular show the average acoplanrity within the ATLAS selection, for the different ZDC classes. As discussed in~\cite{Harland-Lang:2021ysd} and elsewhere, this is particularly sensitive to the peripherality of the interaction, and hence we may expect an impact on the acoplanarity distribution from requiring coincident production. This is indeed observed in the figure, where we can see that by requiring coincident $\rho^0$ production, the average impact parameter is reduced and therefore the average photon transverse momentum, and hence dimuon acoplanarity, is increased. The relative increase for the difference ZDC classes is not uniform, in line with the discussion above.

We next consider a different process, namely dilepton production where no coincident process is required. Of particular note here is the fact that in such measurements an exclusivity requirement is made to select the data. Therefore, any coincident particle production of the type discussed above that does not pass the particular experimental veto will be rejected. The overall expectation would therefore be that the theoretical predictions, where no such veto is imposed, will overshoot the data by the coincidence fraction, if it is not appropriately corrected for this effect. Interestingly there is in some cases  a trend for some overshoot to occur. This is discussed in~\cite{Harland-Lang:2023ohq}, where it is noted that the \texttt{SuperChic} predictions~\cite{Harland-Lang:2018iur} for the ATLAS data on dimuon production in PbPb collisions~\cite{ATLAS:2015wnx,ATLAS:2017sfe} tend to overshoot the data by roughly 10\%. This is clearly larger than the coincident rates for $\rho^0$ production observed in~\cite{ATLAS:2025nac}, but the precise effect will depend on the difference between the event selection applied there, and the (looser) requirement for a veto, where e.g. only one additional track is required to lie in the fiducial region, as well on other coincident processes beyond $\rho^0$ photoproduction that may play a role.

\begin{figure}
\begin{center}
\includegraphics[scale=0.6]{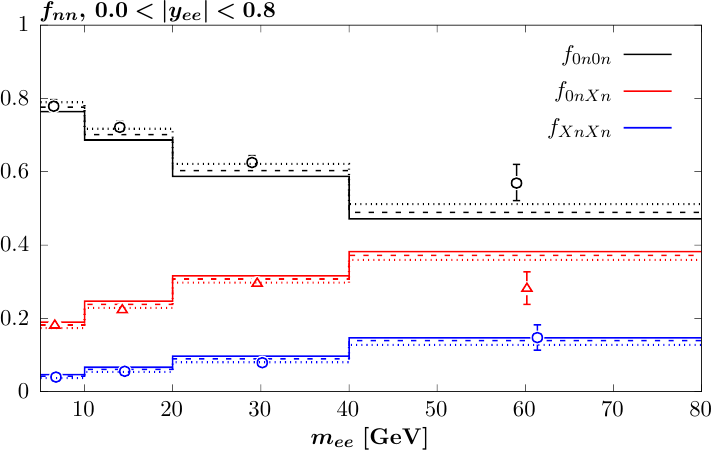}
\includegraphics[scale=0.6]{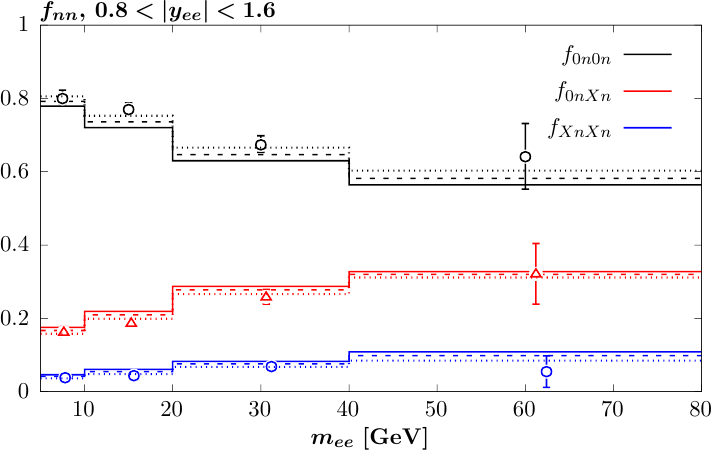}
\caption{\sf Comparison of the \texttt{SuperChic} predictions to the ATLAS data~\cite{ATLAS:2022srr} on ultraperipheral electron pair production in PbPb collisions at $\sqrt{s}_{nn}=5.02$ TeV as a function of the dielectron invariant mass and for different dielectron rapidity bins. Results for the ratio of the $0n0n$, $Xn0n$ and $XnXn$ cross sections to the inclusive UPC case (with respect ion dissociation) are shown. The solid lines correspond to the default predictions, as shown in~\cite{Harland-Lang:2023ohq}, while the dashed (dotted) curves correspond to subtracting the predictions with coincident $\rho^0$ production, multiplied by 5 (10) so as to be visible. The electrons are required to have $p_{\perp, e} > 2.5$ GeV, $|\eta_e| < 2.4$, $m_{ee} > 5$ GeV and $p_{\perp, ee} < 2$ GeV. Data errors correspond to systematic and statistical added in quadrature.
}
\label{fig:ATLASdiel}
\end{center}
\end{figure}

More precisely, it is interesting to note that this reduction is not expected to be uniform with respect to the ZDC selection or kinematics of the primary produced state. For e.g. the $XnXn$ class the coinicidence rate is significantly higher than in the inclusive one, and hence we will expect this veto effect to have some impact on the relative ZDC fractions, as well as the overall rate. This is explored in Fig.~\ref{fig:ATLASdiel}, where a comparison to the ATLAS data~\cite{ATLAS:2022srr} on ultraperipheral electron pair production in PbPb collisions at $\sqrt{s}_{nn}=5.02$ TeV  is presented. The baseline predictions are as shown in~\cite{Harland-Lang:2023ohq}, and if we take these and simply subtract the cross sections with coincident production corresponding to the selection of~\cite{ATLAS:2025nac}, then indeed this percent level effect is hardly visible on the plot. However, it is interesting to note that if we inflate the overall rate by e.g. a factor of 5, then overall trend is to improve the description of the data. It seems likely that this level of inflation is too extreme, but bearing in mind the differences between the selection of~\cite{ATLAS:2025nac} and any veto imposed in this analysis, it remains a potentially relevant observation; probably more relevant is the fact that multiple electron emission will have a similar effect. While we do not consider it explicitly here, similar conclusions will hold for differential distributions, namely the (percent--level) mismatch due to this veto effect will not be constant but will follow the expectations of e.g. Fig.~\ref{fig:dists}.

\section{Conclusions}\label{sec:conc}

In this paper we have considered the case of coincident particle production in  ultraperipheral heavy ion collisions, that is the production of some state of interest in addition to the primary UPC process under study, due to multiple initial--state photon emission. We have focussed on the case of photon--initiated dimuon production with coincident $\rho^0$ meson photoproduction in PbPb collisions, which is included in the \texttt{SuperChic} MC generator, and is publicly available at
\begin{center}
    \href{https://github.com/LucianHL/SuperChic}{https://github.com/LucianHL/SuperChic}.
\end{center}
This is motivated by the recent ATLAS measurement~\cite{ATLAS:2025nac} of this process. We have presented the first precise theoretical calculation of the expected rate for this, and the non--trivial dependence of it on the ZDC selection of the colliding ions as well as the kinematics of the primary dimuon state. Results for this are found to be in very encouraging agreement with the data. The non--trivial dependence of the coincidence rate on the ZDC selection and dimuon kinematics is in particular reproduced well.

In addition to coincident photoproduction, we have also discussed how this approach can be readily extended to the case of coincident two photon--initiated production, in particular of lepton pairs. The general expectations for this are that multiple muon pair production is expected to be rather strongly suppressed, whereas multiple electron pair production close to threshold should be ubiquitous. This is consistent with other results in the literature.

We have also considered the possible impact of such coincident production more generally on standard UPC measurements. In particular, as an exclusivity veto is applied in this cases, any such coincident production that results in additional particles in the veto region will be rejected. If this is not currently corrected for, then given it is not included in existing theoretical calculations, this will result in some overshoot of the theory in comparison to data. As discussed in this paper, due to the kinematic and ZDC class dependence of the coincidence probability, this effect will not be constant but will rather depend on these variables. 

In the future, it will be interesting to extend this study to include other coincident production processes, such as multiple lepton production, which play a key role in the veto efficiency question discussed above. Similarly, the coincident photoproduction of other states ($\omega,\phi, J/\psi$...), while expected to occur with a significantly lower probability, by 1-2 orders of magnitude, may be observable, and the corresponding rates are readily calculable using the same approach as for $\rho^0$ production. More broadly the non--trivial dependence of coincident production on the peripherality of the UPC process provides a novel testing ground with which to probe this production mechanism, with potential implications for a range of studies, form nuclear structure to beyond the Standard Model physics searches. Therefore there is much scope for further study and exploration.

\section*{Acknowledgements}

I thank Valery Khoze and Misha Ryskin for reading the manuscript and providing comments. I thank Daniel Tapia Takaki for bringing to my attention the issue of the non--resonant dipion contribution to the ATLAS data. I thank Brian Cole and Ronan McNulty for useful discussions on this topic, and Soumya Mohapatra for providing the data corresponding to the ATLAS measurement. I thank the Science and Technology Facilities Council (STFC) part of U.K. Research and Innovation for support via the grant award ST/T000856/1.

\bibliography{references}{}

\begin{thebibliography}{10}

\bibitem{ATLAS:2020epq}
ATLAS, G.~Aad {\em et~al.},
\newblock Phys. Rev. C {\bf 104}, 024906 (2021), 2011.12211.

\bibitem{ATLAS:2020hii}
ATLAS, G.~Aad {\em et~al.},
\newblock JHEP {\bf 03}, 243 (2021), 2008.05355,
\newblock [Erratum: JHEP 11, 050 (2021)].

\bibitem{ATLAS:2022srr}
ATLAS, G.~Aad {\em et~al.},
\newblock JHEP {\bf 2306}, 182 (2023), 2207.12781.

\bibitem{ATLAS:2022ryk}
ATLAS, G.~Aad {\em et~al.},
\newblock Phys. Rev. Lett. {\bf 131}, 151802 (2023), 2204.13478.

\bibitem{CMS:2022arf}
CMS, A.~Tumasyan {\em et~al.},
\newblock Phys. Rev. Lett. {\bf 131}, 151803 (2023), 2206.05192.

\bibitem{CMS:2020skx}
CMS, A.~M. Sirunyan {\em et~al.},
\newblock Phys. Rev. Lett. {\bf 127}, 122001 (2021), 2011.05239.

\bibitem{CMS:2024bnt}
CMS, A.~Hayrapetyan {\em et~al.},
\newblock (2024), 2412.15413.

\bibitem{ATLAS:2025nac}
ATLAS, G.~Aad {\em et~al.},
\newblock (2025), 2504.07795.

\bibitem{STAR:2019wlg}
STAR, J.~Adam {\em et~al.},
\newblock Phys. Rev. Lett. {\bf 127}, 052302 (2021), 1910.12400.

\bibitem{Broz:2019kpl}
M.~Broz, J.~G. Contreras, and J.~D. Tapia~Takaki,
\newblock Comput. Phys. Commun. {\bf 253}, 107181 (2020), 1908.08263.

\bibitem{Klein:2020fmr}
S.~Klein and P.~Steinberg,
\newblock Ann. Rev. Nucl. Part. Sci. {\bf 70}, 323 (2020), 2005.01872.

\bibitem{Harland-Lang:2023ohq}
L.~A. Harland-Lang,
\newblock Phys. Rev. D {\bf 107}, 093004 (2023), 2303.04826.

\bibitem{Hencken:2006ir}
K.~Hencken, E.~A. Kuraev, and V.~Serbo,
\newblock Phys. Rev. C {\bf 75}, 034903 (2007), hep-ph/0606069.

\bibitem{Krachkov:2014gba}
P.~A. Krachkov, R.~N. Lee, and A.~I. Milstein,
\newblock Phys. Rev. A {\bf 90}, 062112 (2014), 1410.6566.

\bibitem{Klusek-Gawenda:2016suk}
M.~K\l{}usek-Gawenda and A.~Szczurek,
\newblock Phys. Lett. B {\bf 763}, 416 (2016), 1607.05095.

\bibitem{vanHameren:2017krz}
A.~van Hameren, M.~K\l{}usek-Gawenda, and A.~Szczurek,
\newblock Phys. Lett. B {\bf 776}, 84 (2018), 1708.07742.

\bibitem{Zha:2021jhf}
W.~Zha and Z.~Tang,
\newblock JHEP {\bf 08}, 083 (2021), 2103.04605.

\bibitem{Klein:1999qj}
S.~Klein and J.~Nystrand,
\newblock Phys. Rev. C {\bf 60}, 014903 (1999), hep-ph/9902259.

\bibitem{Harland-Lang:2018iur}
L.~A. Harland-Lang, V.~A. Khoze, and M.~G. Ryskin,
\newblock Eur. Phys. J. C {\bf 79}, 39 (2019), 1810.06567.

\bibitem{Harland-Lang:2019zur}
L.~Harland-Lang, J.~Jaeckel, and M.~Spannowsky,
\newblock Phys. Lett. B {\bf 793}, 281 (2019), 1902.04878.

\bibitem{Budnev:1975poe}
V.~M. Budnev, I.~F. Ginzburg, G.~V. Meledin, and V.~G. Serbo,
\newblock Phys. Rept. {\bf 15}, 181 (1975).

\bibitem{Baur:2001jj}
G.~Baur, K.~Hencken, D.~Trautmann, S.~Sadovsky, and Y.~Kharlov,
\newblock Phys. Rept. {\bf 364}, 359 (2002), hep-ph/0112211.

\bibitem{Azevedo:2023vsz}
C.~N. Azevedo, V.~P. Goncalves, and B.~D. Moreira,
\newblock Eur. Phys. J. A {\bf 59}, 193 (2023), 2306.05519.

\bibitem{Harland-Lang:2011mlc}
L.~A. Harland-Lang, C.~H. Kom, K.~Sakurai, and W.~J. Stirling,
\newblock Eur. Phys. J. C {\bf 72}, 1969 (2012), 1110.4320.

\bibitem{Harland-Lang:2021ysd}
L.~A. Harland-Lang, V.~A. Khoze, and M.~G. Ryskin,
\newblock SciPost Phys. {\bf 11}, 064 (2021), 2104.13392.

\bibitem{Klein:2016yzr}
S.~R. Klein, J.~Nystrand, J.~Seger, Y.~Gorbunov, and J.~Butterworth,
\newblock Comput. Phys. Commun. {\bf 212}, 258 (2017), 1607.03838.

\bibitem{ALICE:2020ugp}
ALICE, S.~Acharya {\em et~al.},
\newblock JHEP {\bf 06}, 035 (2020), 2002.10897.

\bibitem{Frankfurt:2015cwa}
L.~Frankfurt, V.~Guzey, M.~Strikman, and M.~Zhalov,
\newblock Phys. Lett. B {\bf 752}, 51 (2016), 1506.07150.

\bibitem{ZEUS:1997rof}
ZEUS, J.~Breitweg {\em et~al.},
\newblock Eur. Phys. J. C {\bf 2}, 247 (1998), hep-ex/9712020.

\bibitem{ALICE:2015nbw}
ALICE, J.~Adam {\em et~al.},
\newblock JHEP {\bf 09}, 095 (2015), 1503.09177.

\bibitem{STAR:2017enh}
STAR, L.~Adamczyk {\em et~al.},
\newblock Phys. Rev. C {\bf 96}, 054904 (2017), 1702.07705.

\bibitem{LHCb:2025fzk}
LHCb, R.~Aaij {\em et~al.},
\newblock (2025), 2506.06250.

\bibitem{ATLAS:2015wnx}
ATLAS, G.~Aad {\em et~al.},
\newblock Phys. Lett. B {\bf 749}, 242 (2015), 1506.07098.

\bibitem{ATLAS:2017sfe}
ATLAS, M.~Aaboud {\em et~al.},
\newblock Phys. Lett. B {\bf 777}, 303 (2018), 1708.04053.

\end{thebibliography}
\bibliographystyle{h-physrev}

\end{document}